\documentstyle{aipproc}
\begin{document}
\renewcommand{\baselinestretch}{1.2}
\title{Sandpiles and absorbing-state phase transitions:
 \\recent results and open problems}    
 
\def\xv{ {\bf x}}      
\author{
Miguel A. Mu\~noz$^1$, Ronald Dickman $^2$,
Romualdo Pastor-Satorras $^3$,
Alessandro Vespignani $^4$, and Stefano Zapperi $^5$  }
\address{
$^1$  Institute {\it Carlos I} for Theoretical
and Computational Physics\\ and Depto. de E. y
F{\'\i}sica de la Materia, Universidad de Granada,
18071 Granada, Spain.\\  
$^2$ 
  Departamento de F\'{\i}sica, ICEx,
Universidade Federal de Minas Gerais,
Caixa Postal 702,
30161-970 Belo Horizonte, MG, Brazil\\
$^3$ Dept. de F\'{\i}sica i E. Nuclear,
Universitat Polit\`{e}cnica de Catalunya,
Campus Nord, M\'{o}dul B4,  08034 Barcelona, Spain\\ 
$^4$ The Abdus Salam Int. Centre for Theoretical Physics
P.O. Box 586, 34100 Trieste, Italy\\
$^5$ INFM, Dipartimento di Fisica, E. Fermi,
Universit\'a de Roma "La Sapienza", P.le A. Moro 2,
00185 Roma, Italy \\ }


\maketitle

\begin{abstract}
We review some recent results on the relations
between sandpiles and a class of absorbing state phase
transitions. We use the concept of fixed 
energy sandpiles (FES), in which external
 driving and dissipation are absent.
 FES are shown to exhibit an absorbing 
state transition with critical properties 
coinciding with those of the 
corresponding sandpile model.
We propose a set of Langevin equations
capturing the relevant features of this transition.
These equations characterize the universality class of 
systems with an infinite number of absorbing states and 
a static conserved field coupled to the order parameter. 
Different models in this class are identified, and strong 
evidence is presented showing that the Manna sandpile,
as well as some other stochastic sandpiles, belong in this
universality class. Finally some 
open problems and questions are discussed.
\end{abstract}

\section*{Introduction}
 
 Slowly driven by the input of new ideas and concepts,
avalanches of papers have in recent years seen the light 
 in the field of self-organized criticality (SOC)
\cite{btw,manna,zhang1,Dhar,Jensen,2times,stella,BJP}.
  This activity was certainly justified by the huge
conceptual and practical implications 
of the self-organization mechanisms. 
From a theoretical point of view
two main questions were soon posed: (i)
what are the main ingredients able to drive ``spontaneously''
a system to its critical state without need of fine-tuning
any control parameter?, and (ii)
what are the links between self-organized
critical states and standard critical ones appearing 
in equilibrium and non-equilibrium systems?.   
In order to answer the first of the two aforementioned
issues, several paths to SOC have been identified (a 
recent review of them, as well as a summary on experiments on SOC
can be found in \cite{BJP}). 
Certainly the most celebrated ones among them are:
the presence of two infinitely-separated time scales
\cite{2times},
and extremal dynamics \cite{bs}, though
these are certainly not the only ones \cite{Dhar,BJP}.  

 For example, in sandpiles \cite{btw,manna,zhang1},
the archetype of SOC systems,
the two infinitely separated time scales correspond to
a slow scale at which grains of sand are added to the system,
and a fast one for the relaxation of unstable piles which leads 
to a rapid redistribution of sand, and eventually to 
boundary dissipation.
 It has been shown that it is only in the double 
limit of driving rate, $h$,
and dissipation rate, $\epsilon$ going to $0^+$ with
$h/\epsilon \rightarrow 0$, that scale invariant 
behavior is generated up to the system-size cut-off
 \cite{2times,vz}. Under the effect of the external drive,
the system jumps among stable (absorbing) configurations via
avalanche-like rearrangements. Some of the sandpiles
 we will refer to are: the original Bak-Tang-Wiesenfeld (BTW)
sandpile \cite{btw}, the Manna model \cite{manna},
the Zhang model \cite{zhang1}, the Oslo ricepile model   
\cite{oslo}, as well as some variations of them.

Despite of some very remarkable rigorous results
\cite{Dhar,priezz}
 and some renormalization group analyses \cite{rg},
the question of whether the critical properties of sandpiles 
can be related to more standard critical phase transitions,
 or more specifically, whether the associated critical (avalanche)
exponents \cite{btw,manna,Jensen}
can be ascribed to any known universality class, 
remains open.

 A priori, these are not straightforward questions to answer;
in particular, the presence of two time
scales makes it difficult to apply standard techniques for the
study of scale invariant systems to sandpiles.
 Another intrinsic difficulty lies in the fact, that
exponents controlling the deviation from criticality cannot
be measured in slowly driven sandpiles, posed by definition
at their critical point. 

\section*{Fixed Energy Sandpiles and absorbing states}
  In order to circumvent these difficulties we
have  studied the
 the so-called {\em fixed energy sandpiles} (FES)
\cite{fes2,dvz,FES}. 
These are cellular automata sharing the same microscopic dynamics
as their corresponding standard sandpile
counterparts but without external driving or dissipation.  
For each standard sandpile model 
we can define its FES version by keeping constant 
its total energy (number of sand grains).
In this way, the initial energy density 
$\zeta$ becomes a control parameter. FES are
found to be critical only for a particular value 
$\zeta=\zeta_c$ (which as we will show
 turns out to be identical 
to the stationary energy density of its
driven-dissipative counterpart \cite{FES}).
For energy densities below $\zeta_c$ the FES-sandpile
 falls into an {\it absorbing state}
\cite{reviews}, while for values above that threshold activity 
is indefinitely sustained, i.e. the FES is in
an {\it active phase}. 
Consequently, the phase transition in FES
can be enviewed as an absorbing state phase transition
\cite{reviews,ant} (with an infinite number of absorbing
 states). Observe that, switching on dissipation and driving,
in the presence of activity $\zeta > \zeta_c$ and $d \zeta/dt < 0$.
In the absence of activity there is addition, but no loss of activity,
so $\zeta < \zeta_c$ implies $d \zeta /dt >0$  \cite{inf}.
 Evidently, the only possible stationary
value for the density in the sandpile is $\zeta_c$!
Not only $\zeta_c$ is common to a critical FES and its
slowly-driven counterpart: also higher moments and
correlation functions coincide in the large-size limit
 (as expected, given that the bulk dynamics is identical
 in both cases).   
We have performed extensive numerical analyses 
 of the critical behavior of FES aimed at
 verifying the previous arguments and at
determining a complete set of exponents. We refer
the reader to \cite{FES} where the main ideas, simulations,
analysis techniques, and results are presented
(some of the obtained critical exponents
 are included in table 1).  

 Let us remark that other attempts have been made 
to connect sandpiles with different 
types of phase transitions, as for instance: 
pinning-depinning transitions in disordered media
 \cite{interfaces}, the voter model \cite{directed},
and branching processes \cite{branch}.  It is beyond
the scope of this paper to review or discuss them.
  
  Generically systems exhibiting a transition into an
absorbing state belong in the universality class of directed
 percolation (DP), the field theoretical (or Langevin equation) 
representation of which is the celebrated Reggeon field theory 
(RFT) \cite{rft,reviews}.
 This universality class has proven very
robust with respect to the variation of microscopic details, 
number of components, and number of absorbing states
\cite{pcp,many,reviews}. 
Only in the presence of some extra conservation law or additional
symmetry (as, for example, a $Z_2$-symmetry between two
equivalent absorbing states, which in some cases tantamount to 
{\em parity conservation} \cite{baw}), have 
the critical exponents  been found 
to differ from those of DP \cite{baw}.

Following rather general principles of relevance and symmetry, 
a set of Langevin equations aimed at capturing the critical
features of FES was proposed in \cite{FES}. 
Let us emphasize that this type of phenomenological approach
has proven extremely useful, and provided sound 
results when applied to other systems 
with many absorbing states (see \cite{many}).
This set of equations includes one 
for the evolution of a coarse-grained activity field, 
$\phi(\xv,t)$, and one for 
the energy (or background) field, $E(\xv,t)$ \cite{FES}.
The activity field represents the local density of unstable
sites, while the energy field stands for 
the total number of grains regardless of the local
activity state.
The proposed set of equations is:
\begin{eqnarray}
\partial_t \phi(\xv,t) &= & \mu \phi(\xv,t) - b \phi^2(\xv,t)
 + D \nabla^2 \phi(\xv,t) +  w E(\xv,t) \phi(\xv,t)+ \sigma
 \sqrt{\phi} \eta(\xv,t) 
\nonumber \\
\partial_t E(\xv,t) &= & \lambda \nabla^2 \phi(\xv,t)
\label{Lan}
\end{eqnarray}
where $\mu, b, D, w, \sigma$, and $\lambda$ are constants, and
$\eta$ is a Gaussian white noise.  
In a nutshell, these equations express the fact that the 
transition for the activity is controlled by the same type
of terms appearing in the RFT, i.e. the most relevant
terms in standard absorbing-state phase transitions 
\cite{reviews,rft}, plus an
additional coupling between the activity
field and a static conserved energy field. This extra
term stems from the fact that creation of activity 
is locally fostered by the presence of a high density 
of the background field, and this background field (energy)
is a conserved one. The extra conservation law is therefore 
the new relevant ingredient with respect to RFT.
Some other terms, consistent with the 
symmetries and conservation laws, could have been included
in Eq.(\ref{Lan}) but they all turn out to be irrelevant from a
power counting analysis \cite{FES}. 
Let us also remark that the second equation in  Eq.(\ref{Lan}),
being linear, can be integrated out, leading to a non-Markovian
contribution and a quenched disordered linear term in 
the activity equation 
(a detailed analysis and discussion of 
these terms can be found in \cite{FES}).
In what follows, we avoid performing such an integration
to make explicit the presence of a conservation law,
and stress that this extra contribution is relevant  
at the RFT fixed point \cite{FES}.

 Which sandpiles is this theory intended to describe?
Although in principle we would like to describe the original
BTW sandpile this seems to be an unrealizable task,
 the reasons for this being manifold. The main reason, 
is that the BTW model being deterministic, has 
many hidden conservation laws and toppling invariants \cite{Dhar}
 which are almost impossible to implement in a coarse grained 
Langevin equation (for which a detailed book-keeping of symmetries
and conservation laws is essential). Owing to the existence
of these  conservation laws, the FES version of
the BTW exhibits strong non-ergodicity as shown and discussed
in \cite{FES}; the stationary state depends strongly upon
the initial configuration (and not only on the initial energy).
From a different perspective, it has also been recently 
shown that BTW does not obey simple scaling \cite{stella}.

   Our approach is therefore intended to reproduce the 
scaling behavior of stochastic sandpiles 
with no extra conservation, symmetry, 
or spatial anisotropy \cite{directed}, as for example: 
the Manna model \cite{manna}, ricepiles \cite{oslo}, the
stochastic version of the Zhang model \cite{zhang1,romu3}, and
other modifications of them.

\section*{The universality class}

Having proposed a set of Langevin equations for stochastic 
sandpiles, it would be an extremely desirable next step
to derive specific predictions (i.e. values of the critical
exponents and scaling functions), obtained after solving 
perturbatively the renormalization group associated 
with Eq.(\ref{Lan}).
 Despite considerable efforts, unfortunately, so far we have not 
been able to complete satisfactorily such a challenging 
task. Elucidating whether this is just due to technicalities,
or has some more profound origin (concerning the theory 
renormalizability) still needs to be sorted out.   
  
    \renewcommand{\baselinestretch}{1.2}

\begin{table}
\begin{tabular}{lcccccc}
  &  \multicolumn{6}{c}{Steady state exponents $d=2$} \\
  \cline{2-7}
  & $\beta$ & $\nu_\perp$ & $\beta/\nu_\perp$ & $z$ & & $\theta$\\
  \hline
  CTTP  &$0.64(1)$& $0.82(3)$  & $0.78(3)$ & $1.55(5)$ & & $0.43(1)$\\
  CRD   &$ 0.65(1)$ & $0.83(3)$ & $0.78(2)$ & $1.55(5)$ & & $0.49(1)$\\
  Manna & $0.64(1)$ & $0.82(3)$ & $0.78(2)$ & $1.57(4)$ &  &$0.42(1)$ \\ 
  DP & $0.583(4)$ & $0.733(4)$ & $0.80(1)$ & $1.766(2)$ & & $0.451(1)$ \\
  \hline \hline
  &  \multicolumn{6}{c}{Spreading exponents $d=2$} \\
  \cline{2-7}
  & $\tau_s$  & $D$ & $\tau_t$ & $z$ & $\eta$  &  $\delta$ \\
  \hline
  CTTP &  $1.28(1)$ & $2.76(1)$ & $1.49(1)$ & $1.54(2)$ & $0.29(5)$ &
  $0.50(2)$ \\
  CRD  & $1.28(1)$ & $2.75(1)$ & $1.50(2)$ & $1.54(2)$ & $0.29(2)$ &
  $0.50(2)$ \\
  Manna & $1.28(1)$ & $2.76(1)$ & $1.48(2)$ & $1.55(1)$ & $0.30(3)$
  & $0.48(2)$ \\  
  DP    & $1.268(1)$ & $2.968(1)$ & $1.450(1)$ & $1.766(2)$ & $0.230(1)$ &
  $0.451(1)$
\end{tabular}
\vspace*{0.5cm} 
\begin{tabular}{lcccccc}
  &  \multicolumn{6}{c}{Steady state exponents $d=3$} \\
  \cline{2-7}
  & $\beta$ & $\nu_\perp$ & $\beta/\nu_\perp$ & $z$ & & \\
  \hline
  CRD &$0.86(2)$  & $0.63(5)$ & $1.39(4)$ &$1.80(5)$ &$$\\
  Manna & $0.84(2)$ & $0.60(3)$ & $1.40(2)$& $1.80(5)$ \\ 
  DP & $0.81(1)$ & $0.581(5)$ & $1.39(3)$ & $1.901(5)$ & &  \\
  \hline \hline
  &  \multicolumn{6}{c}{Spreading exponents $d=3$} \\
  \cline{2-7}
  & $\tau_s$  & $D$ & $\tau_t$ & $z$ & $\eta$  &  $\delta$ \\
  \hline
  CRD  & $1.42(1)$ & $3.36(1)$ & $1.80(2)$ & $1.77(1)$ & $0.16(1)$ &
  $0.76(1)$ \\
  Manna & $1.41(1)$ & $3.36(1)$ & $1.78(2)$ & $1.76(2)$ & $0.16(5)$
  & $0.78(2)$ \\  
  DP    & $1.359(1)$ & $3.507(1)$ & $1.730(1)$ & $1.901(5)$ & $0.114(4)$ &
  $0.730(4)$
\end{tabular}
\caption{Critical exponents for spre\-ading and ste\-ady state
  experiments in $d=2$ and $d=3$. Models: CTTP: Conserved threshold
  transfer process; CRD: Conserved reaction-diffusion model; Manna:
  Abelian Manna sandpile; DP: Directed percolation 
(for the sake of comparison).} 
\end{table}

  Leaving aside for the moment field theoretical 
analysis of Eq.(\ref{Lan}),  
we proceed with a slightly more indirect verification of the
validity of our theory. Inspired by the previous set of equations 
Eq.(\ref{Lan}) two of us have recently made the following  
conjecture:
 {\em all stochastic models with an infinite number of absorbing states 
in which the order parameter evolution is coupled to a
 static conserved field define a unique universality class}
\cite{romu1}. This conjecture has been supported  by extensive 
simulations of the following different models 
fulfilling all the above-listed requirements \cite{romu1,romu2}:
\begin{itemize}
\item (i) a conserved threshold transfer process \cite{cttp,reviews};
\item (ii) a conserved lattice 
gas with repulsion of nearest neighbor
 particles \cite{romu1}; 
\item (iii) a reaction-diffusion model (CRD) with two species
A and B \cite{romu2}
\end{itemize}
(see \cite{romu1,romu2} for model definitions and details). 
 All the measured critical exponents both for steady state
and for analysis of spreading \cite{reviews} coincide
within error bars and are also the same as those measured in
extensive numerical simulations of the FES Manna sandpile
\cite{FES} (up-to-date exponent values in two and three
dimensions are listed in table I; for further
simulation details as well as model and exponents 
definitions see \cite{FES,romu1,romu2,reviews}). 

Furthermore, the slowly-driven versions of the previous models 
are self-organized \cite{romu1,romu2},  
and the corresponding avalanche exponents are in good agreement
with those of stochastic sandpiles \cite{Num-manna,FES}. 
Finally, all the scaling laws among 
steady-state, spreading and avalanche critical exponents
\cite{Brief} are satisfied 
within numerical accuracy (except for some well identified 
anomalies \cite{FES,romu1,romu2}). 
This shows rather unambiguously that the critical point of 
a slowly-driven sandpile coincides
with that of its FES version, supports
strongly the previously enunciated conjecture, and identifies
stochastic sandpiles as belonging in the universality class
of systems with many absorbing states and order parameter
coupled to a static conserved field.

 Moreover, the conserved reaction-diffusion model studied
in \cite{romu2} has the notorious advantage of been 
susceptible to be mapped exactly using a a Fock-space
representation and creation-annihilation operators 
\cite{Fock} into an effective action, or equivalently 
into a set of Langevin equations \cite{romu2}. 
Remarkably, up to na{\"\i}vely irrelevant terms, this set
of equations coincides with Eq.(\ref{Lan}). The additional
irrelevant terms appearing in the exact mapping,
might also be of some importance in lower dimensions
(this issue is currently under study).

\section*{Conclusions and open problems}

  Different stochastic sandpile models show critical properties
that are fully compatible with those of their
corresponding FES versions. Using this equivalence
stochastic
sandpiles have been shown to belong in the universality class
of systems exhibiting an absorbing-state phase transition
 with order parameter 
coupled to a static conserved density field.
This result places sandpiles within the range of 
applicability of standard tools for the study 
of scale invariance in phase transitions, and constitutes in our
opinion a step forward in the rationalizing SOC.

 Finally let us enumerate some open questions 
and problems, on which we are presently working:
\begin{itemize} 
\item Direct numerical integration of the Langevin equations
Eq.(\ref{Lan}) using the method of \cite{discrete}. This 
will provide an even more direct test for the validity
of the Langevin equations.
\item Connection of the universality class
discussed in this paper with that of the 
pinning-depinning transitions in disordered systems
\cite{interfaces}.
Numerical simulations show that these two
classes are indeed the same; we
are working on the theoretical connections between 
them
 \cite{AM}.
\item Renormalization of the field theory for FES, Eq. (\ref{Lan}).
\item Clarification of some anomalies found in some time-dependent
magnitudes in FES (like that affecting the exponent $\theta$).
\item Extension of the numerical analysis 
to one-dimensional sandpiles.
\end{itemize}

\vspace{0.3cm}
 
{\bf ACKNOWLEDGMENTS}
We acknowledge M. Alava, A. Barrat, D. Dhar, 
P. Grassberger, K.B. Lauritsen, 
E.Marinari, L. Pietronero and A. Stella
for useful discussions and comments.
We acknowledge partial support
from the European Network contract ERBFMRXCT980183;
M.A.M acknowledges also support from the 
Spanish DGESIC project PB97-0842, and
Junta de Andaluc{\'\i}a project FQM-165.  
R.D. acknowledges CNPq and CAPES.
R.P.R. acknowledges support from grant CICYT PB97-0693.



\begin{references}
     
\bibitem{btw}
        P. Bak, C. Tang and K. Wiesenfeld,
        Phys. Rev. Lett. {\bf 59}, 381 (1987);
        Phys. Rev. A {\bf 38}, 364 (1988).
                   
\bibitem{manna}
        S. S. Manna,
        J. Phys. A {\bf 24}, L363 (1991).

\bibitem{zhang1}
 Y.-C. Zhang, Phys. Rev. Lett. {\bf 63}, 470 (1989).
     
\bibitem{Dhar}
D. Dhar, Phys. Rev. Lett.{\bf 64}, 1613 (1990);
Physica A {\bf 263}, 4 (1999).
S. N. Majumdar and D. Dhar, Physica A {\bf 185},
 129 (1992). For recent reviews see
also D. Dhar, Physica A {\bf 264}, 1 (1999); and  cond-mat/9909009.

\bibitem{Jensen} H. J. Jensen, {\it Self organized criticality},
(Cambridge Univ. Press, Cambridge, 1998).

\bibitem{2times}
G. Grinstein, in
{\it Scale Invariance, Interfaces and Nonequilibrium Dynamics},
{\it NATO Advanced Study Institute, Series B: Physics},
vol. 344, A. McKane et al., Eds.
(Plenum, New York, 1995). T. Hwa and M. Kardar,
Phys. Rev. A {\bf 45}, 7002 (1992).
See also    A. Barrat, A. Vespignani and S. Zapperi,
Phys. Rev. Lett. {\bf 83}, 1962 (1999).

\bibitem{stella}
M. De Menech, A. L. Stella and C. Tebaldi,
Phys. Rev. E {\bf 58}, R2677 (1998);
C. Tebaldi, et al., Phys. Rev. Lett. {\bf 83}, 3952 (1999).
M. De Menech, A. L. Stella, Phys. Rev. E {\bf 62}, R4528 (2000)
See also; D. V. Ktitarev et al. Phys. Rev. E
{\bf 61}, 81 (2000). 

\bibitem{BJP} R. Dickman, M. A. Mu\~noz, A. Vespignani,
and S. Zapperi, Braz. J. of Phys. {\bf 30}, 27 (2000).

\bibitem{bs}
P. Bak and K. Sneppen,
Phys. Rev. Lett. {\bf 71}, 4083 (1993). 
K. Sneppen, Phys. Rev. Lett. {\bf 62}, 3539 (1992).
A. Hansen and S. Roux, J. Phys A {\bf 20}, L873 (1987).
P. Grassberger and Y.-C. Zhang, Physica A {\bf 224}, 169 (1996).
Z. Olami, I. Procaccia, and R. Zeitak, Phys. Rev. E {\bf 49},
1232 (1994).  S. Maslov and Y-C. Zhang, Phys. Rev. Lett. {\bf
75}, 1550 (1995).
F. Bagnoli et al., Phys. Rev. E {\bf 55}, 3970 (1997).  
 
\bibitem{vz}
A. Vespignani and S. Zapperi,
Phys. Rev. Lett. {\bf 78}, 4793 (1997);
Phys. Rev. E {\bf 57}, 6345  (1998).
 
\bibitem{oslo}  K. Chistensen, et al. Nature (London)
{\bf 379}, 49 (1996). L.A.N. Amaral and K. B. Lauritsen,
Phys. Rev. E {\bf 54}, 4512 (1996).  


\bibitem{priezz}V. B. Priezzhev, J. Stat. Phys. {\bf 74}, 955 (1994);
E. V. Ivashkevich, J. Phys. A {\bf 27}, 3643 (1994);
 E. V. Ivashkevich, et al.,
Physica A {\bf 209}, 347 (1994). V. B. Priezzhev, cond-mat/9904054.

 
\bibitem{rg}
 A. D{\'\i}az-Guilera, Europhys. Lett. {\bf 26}, 177 (1994). 
L. Pietronero, A. Vespignani and S. Zapperi,
Phys. Rev. Lett. {\bf 72}, 1690 (1994).
J. Hasty and K. Wiesenfeld, J. Stat. Phys. {\bf 86}, 1179 (1997).

\bibitem{fes2} C. Tang and P. Bak, Phys. Rev. Lett.
{\bf 60}, 2347 (1988). See also,
A. Montakhab and J. M. Carlson,
Phys. Rev. E {\bf 58}, 5608 (1998).    

\bibitem{dvz}
        R. Dickman, A. Vespignani and S. Zapperi,
        Phys. Rev. E {\bf 57}, 5095 (1998). 


\bibitem{FES} A. Vespignani, R. Dickman, 
M. A. Mu\~noz, and S. Zapperi,
Phys. Rev. Lett. {\bf 81}, 5676 (1998); 
Phys. Rev. E {\bf 62}, 4564 (2000).
 


\bibitem{reviews}
 R. Dickman in {\em Nonequilibrium Statistical Mechanics in One Dimension}
V. Privman, Ed. (Cambridge University Press, Cambridge 1996);
G. Grinstein and M. A. Mu\~noz, in
 {\it Fourth Granada Lectures in Computational} 
Ed. P. Garrido and J. Marro,
Lecture Notes in Physics, {\bf 493}, 223
(Springer-Verlag, Berlin, 1997).
J. Marro and R. Dickman,
{\em Nonequilibrium Phase Transitions in Lattice Models}
(Cambridge University Press, Cambridge, 1999).
H. Hinrichsen, 
{\it Nonequilibrium Critical Phenomena and Phase
 Transitions into Absorbing States}, cond-mat/0001070.

 
\bibitem{inf}
Like any other statistical model, a fixed-energy sandpile exhibits
critical singularities only in the infinite-size limit.  In this limit
the activity density is strictly zero for $\zeta < \zeta_c$, and
positive for $\zeta > \zeta_c$, ensuring the stated inequality
for $d \zeta/dt$ in the slowly-driven system.    

                                                     
\bibitem{ant}
Connections between absorbing state phase transitions
 and SOC models have been proposed in the past:
M. Paczuski, S. Maslov, and P. Bak,
Europhys. Lett. {\bf 27}, 97 (1994).
S. Maslov and Y-C. Zhang, Physica A {\bf 223}, 1 (1996).
D. Sornett andI. Dornic, Phys. Rev. E {\bf 54}, 3334 (1996).
D. Sornette, et al., J. Phys. I (France) {\bf 5}, 325 (1995).
P. Grassberger, Phys. Lett. A{\bf 200}, 277 (1995).        
A. V\'azquez and O. Sotolongo-Costa, J. Phys. A {\bf 
32}, 2633 (1999); Phys. Rev. E {\bf 61}, 944 (2000).


\bibitem{interfaces} O. Narayan and A. A. Middleton,
 Phys. Rev. B {\bf 49} 244 (1994). 
M. Paczuski and S. Boettcher,
Phys. Rev. Lett. {\bf 77}, 111 (1996).
K. B. Lauritsen and M. Alava, cond-mat/9903346. 
M. Alava and K. B. Lauritsen, cond-mat/9902406.

     
\bibitem{directed}
D. Dhar and R. Ramaswamy, Phys. Rev. Lett.
{\bf 63}, 1659 (1989).
B. Tadic and D. Dhar,
Phys. Rev. Lett. {\bf 79}, 1519 (1997).


\bibitem{branch} 
 S. Zapperi, K. B. Lauritsen, and H. E. Stanley,
Phys. Rev. Lett. {\bf 75}, 4071 (1995).
E. V. Ivashkevich, Phys. Rev. Lett. {\bf 76}, 3368 (1996). 

\bibitem{rft}
  J.L. Cardy and R.L. Sugar, J. Phys. A  {\bf 13}, L423 (1980);
  P. Grassberger, Z. Phys. B {\bf 47}, 365 (1982);
  H.K. Janssen, Z. Phys. B {\bf 42}, 151 (1981).


\bibitem{pcp}
I. Jensen, Phys. Rev. Lett. {\bf 70}, 1465 (1993);
I. Jensen and R. Dickman, Phys. Rev. E {\bf 48}, 1710 (1993).

\bibitem{many}  M. A. Mu\~noz, G. Grinstein, R. Dickman 
and R. Livi, Phys. Rev. Lett. {\bf 76}, 451, (1996).
M. A. Mu\~noz, et al.,
J. Stat. Phys. {\bf 91}, 541-569 (1998).

\bibitem{baw}See:
        P. Grassberger et al.,
        J. Phys. A {\bf 17}, L105 (1984);
        N. Menyhard and G. \'Odor,
        J. Phys. A {\bf 29}, 7739 (1996);
        J. Cardy and U. C. T\"auber,
        Phys. Rev. Lett. {\bf 77}, 4780 (1996);
        H. Hinrichsen,
        Phys. Rev. E {\bf 55}, 219 (1997);
       W. Hwang, S. Kwon, H. Park, and H. Park,
        Phys. Rev. E {\bf 57}, 6438 (1998);
 and references therein.   


\bibitem{romu3}   R. Pastor-Satorras and A. Vespignani, 
cond-mat/0010223. In press (Eur. Phys. J. B.)


\bibitem{romu1} M. Rossi, R. Pastor-Satorras and A. Vespignani,
       Phys. Rev. Lett. {\bf 85}, 1803 (2000). 
 
\bibitem{romu2}
        R. Pastor-Satorras and A. Vespignani,
       Phys. Rev. E{\bf 62 }, R5875 (2000). The model studied
in this reference is a limiting case of a more general class
studied in F. van Wijland, K. Oerding, and H. J. Hilhorst, 
Physica A{\bf 251 }, 179 (1998).

\bibitem{cttp} This is a conserved counterpart of the model introduced
in: J.F.F Mendes et al.
J. Phys. A {\bf 27}, 3019 (1994).           

\bibitem{Num-manna}
        P. Grassberger and S. S. Manna,
        J. Phys. (France) {\bf 51}, 1077 (1990).
        S. S. Manna,
        J. Stat. Phys. {\bf 59}, 509 (1990).
        S. L\"ubeck and K.D. Usadel,
        Phys. Rev. E {\bf 55}, 4095 (1997);
        ibid. {\bf 56}, 5138 (1997).

\bibitem{Brief} See, for instance,
 M. A. Mu\~noz,  R. Dickman, A. Vespignani, and S. Zapperi,
Phys. Rev. E {\bf 59}, 6175 (1999); and references therein.
 

\bibitem{Fock} M. Doi, J. Phys. A{\bf 9}, 1465 (1976).
     L. Peliti, J. Physique {\bf 46}, 1469 (1985).
     B. P. Lee and J. Cardy,
     J. Stat. Phys. A {\bf 80}, 971 (1995).                              

\bibitem{discrete} R. Dickman, Phys. Rev. E {\bf 50}, 4404 (1994).
C. L\'opez and M. A. Mu\~noz, Phys. Rev. E {\bf 56}, 4864 (1997).

\bibitem{AM} M. Alava and M. A. Mu\~noz, In preparation.
                            


\end{references}
\end{document}